\documentclass[conference,usletter]{IEEEtran}
\usepackage[left=.80in, right=.80in, bottom=.80in, top=0.80in]{geometry}
\usepackage{mathtools}

\usepackage{amsmath,amssymb,threeparttable,placeins}
\usepackage{lipsum}
\usepackage{bm}
\usepackage{multicol}
\usepackage{array} 

\usepackage[mathscr]{euscript}
\usepackage[dvips]{graphicx}
\usepackage[usenames]{color}
\usepackage{amsfonts}
\usepackage{latexsym}
\usepackage{subfigure}
\usepackage[format=hang]{subfig}
\usepackage[justification=centering]{caption}
\usepackage{floatrow}
\usepackage{multirow}
\usepackage{graphicx}
\usepackage{epstopdf}
\usepackage[english]{babel}
\DeclareMathAlphabet{\mathpzc}{OT1}{pzc}{m}{it}

\newtheorem{theorem}{{{\textit{Theorem}}}}

\newtheorem{definition}{\textit{Definition}}
\newtheorem{proposition}{{{\textit{Proposition}}}}
\newtheorem{remark}{{{ \textit{Remark}}}}

\newtheorem{example}{{{\textit{Example}}}}

\begin{document}

\title{Two-Dimensional $Z$-Complementary Array Quads with Low Column Sequence PMEPRs } 

%%%%%%
\author{\IEEEauthorblockN{{\bf Shibsankar Das}$^{\ast}$, {\bf Adrish Banerjee}$^{\ast}$ and {\bf Udaya Parampalli}$^{\dagger}$}
\vspace{0.2cm}
	\IEEEauthorblockA{$^{\ast}$Department of Electrical Engineering, Indian Institute of Technology Kanpur, Kanpur, India\\
		$^{\dagger}$School of Computing and Information Systems, University of Melbourne, Victoria, Australia\\
		Email:\{shibsankar, adrish\}@iitk.ac.in, udaya@unimelb.edu.au.
	}
}

\maketitle

\begin{abstract}
	In this paper, we first propose a new design strategy of 2D $Z$-complementary array quads (2D-ZCAQs) with feasible array sizes. A 2D-ZCAQ consists of four distinct unimodular arrays satisfying zero 2D auto-correlation sums for non-trivial 2D time-shifts within certain zone. Then, we obtain the upper bounds on the column sequence peak-to-mean envelope power ratio (PMEPR) of the constructed 2D-ZCAQs by using specific auto-correlation properties of some seed sequences. The constructed 2D-ZCAQs with bounded column sequence PMEPR  can be used as a potential alternative to 2D Golay complementary array sets for practical applications.
\end{abstract}

\section{Introduction} 
\label{sec:introduction}
%\subsection{\bf Background and Literature Review}
A Golay complementary pair (GCP) represents a pair of sequences satisfying zero out-of-phase auto-correlation sums \cite{1961golay}. Due to the preferable correlation property, GCPs have found numerous engineering applications, including peak-to-mean envelope power ratio (PMEPR) control in coded orthogonal frequency-division multiplexing (OFDM) systems \cite{1999davis}, \cite{2000Paterson}, and channel estimation \cite{2001Spasojevic}. There have been many construction methods for binary and complex GCPs with various sequence lengths \cite{1999davis}, \cite{1990Budishin}. To obtain the admissible sequence lengths of GCPs, Fan \textit{et al.} proposed the concept of $Z$-complementary pairs (ZCPs) in \cite{2007Fan_ZCSS}. A ZCP satisfies the zero auto-correlation sums within a zone around the zero time-shift. This region is said to be a zero-correlation zone (ZCZ). In \cite{2011LiExiztenceZCP}, it has been proved that binary ZCPs can exist for all lengths. Such GCPs and ZCPs fall into the categories of one-dimensional GCPs (1D-GCPs) and 1D-ZCPs, respectively.  There have been many research activities to design complementary (and $Z$-complementary) sequences    \cite{2014Liu_TIT}\nocite{2014Liu_Z_CCS}\nocite{2017Chen}\nocite{2017APCCShibsankar}\nocite{2017Shibsankar}\nocite{2018Xie}\nocite{2018AvikSPL}\nocite{2018Shibsankar}\nocite{2019ShibsankarTSP}\nocite{2019Chen}\nocite{2019ShibIWSDA}\nocite{2020Shibsankar_TSP2020_II}\nocite{2022ISITShib}-\cite{2020PaiCOML}. 

The idea of 1D-GCPs is extended to 2D array sets, called 2D Golay complementary array sets (2D-GCASs) \cite{2021LiTSP}\nocite{2021WangTIT}-\cite{2022PaiTCOM}. A 2D Golay complementary array quad (2D-GCAQ) refers to a 2D-GCAS containing four distinct 2D arrays and satisfying zero 2D auto-correlation sums for any non-trivial 2D time-shifts. An array quad is a set of four distinct arrays with equal size. Since signal processing of 2D-GCASs uses exponential smoothing, there is a considerable unbalance among distinct phases when the number of phases increases \cite{1980Frank}. Thus, due to the smaller alphabet size, 2D-GCAQs with four-phases can be used to distinguish the signal levels at the receiver \cite{1980Frank}, \cite{2021LiuThree}. However, 2D-GCAQs can exist only for limited array sizes. For instance, a $4$-ary 2D-GCAQ with array size $7\times 3$ may not exist due to the unavailability of $2$-ary and $4$-ary 1D-GCP of length $7$ \cite{2002Craigen}. This motivates us to investigate 2D $Z$-complementary array quads (2D-ZCAQs) with varieties of array sizes and ZCZ widths. 

Owing to their perfect auto-correlation properties, 2D-GCASs and 2D-ZCAQs can also be employed as spreading sequences in the ultra wide band (UWB) multi-carrier CDMA (MC-CDMA) systems \cite{2003ZhangCUWBST}\nocite{2004Turccsany}\nocite{2003Farkas}-\cite{2006Turcsany}. In a 2D-ZCAQ-based UWB MC-CDMA system, the data bits can be transmitted over the time slots by adopting the ``shift and add" operation to obtain an interference-free communication. At the same time, a high PMEPR issue exists as it is based on the OFDM technique. It has been realized that the PMEPR value is connected only with the column sequences of 2D-ZCAQ/2D-GCAS \cite{2003ZhangCUWBST}. In the second dimension (i.e., along $y$-axis), the column sequences of 2D-ZCAQ can take the role of controlling the PMEPR value. That is, 2D-ZCAQs can offer bandwidth efficiency and reduced PMEPR value along $x$-axis and $y$-axis, respectively, in the UWB MC-CDMA system \cite{2003ZhangCUWBST}. Thus, it is important to construct 2D-ZCAQs with substantially lower column sequence PMEPR values. So far, most of the works have studied only the design of 2D $Z$-complementary array pairs (2D-ZCAPs) \cite{2011LiIWSDA}\nocite{2021PaiTIT}\nocite{2021AbhishekCOML}\nocite{2021PaiSPL}\nocite{2020DasTSP2D}-\cite{2021Zhang}. A 2D-ZCAP consists of two distinct arrays satisfying zero 2D auto-correlation sums within the 2D-ZCZ width \cite{2021PaiTIT}. 

In this paper, the main contributions are as follows.
\begin{itemize}
	\item We present a novel design strategy of 2D-ZCAQs with array size $L\times N$ and 2D-ZCZ width  $Z\times N$ by using an arbitrary 1D-GCP of length $N$ and a 1D $(L,Z)$-ZCPs. The proposed construction method is based on generating 2D  polynomials in $z$-domain. We show that the flexible sequence lengths of complex 1D-GCPs and 1D-ZCPs offer many new array sizes and ZCZ widths of 2D-ZCAQs. 
	\item We derive the upper bounds on the column sequence PMEPR of the constructed 2D-ZCAQs by using specific auto-correlation properties of some existing seed 1D-ZCPs.  
	\item We show that the proposed 2D-ZCAQ has the column sequence PMEPRs at most $3.33,\ 4.00$ and $3.714$ when we consider the seed 1D-ZCPs from \cite{2014Liu_Z_CCS}, \cite{2018AvikSPL} and \cite[Th. 1]{2018Xie}, respectively.  
\end{itemize}

%\subsection{\bf Organization}
% The remainder of the paper is organized as follows. In Section \ref{sec:background}, we present some basic definitions and notations. In section \ref{sec:prop}, we propose a new construction method of 2D-ZCAQs with varieties of array sizes. In Section \ref{sec:PMEPR:2D-ZCAQs}, we study the column sequence PMEPR values of our proposed 2D-ZCAQs. The column sequence PMEPR value of the constructed 2D-ZCAQ is controlled by the specific auto-correlation properties of the seed 1D-ZCPs. Finally, we summarize our works in Section \ref{sec:conclusion}. 

\section{Background} 
\label{sec:background}
In this section, basic definitions and notations have been presented.

A 2D array $\hat{\textit{\textbf{X}}}=[\hat{x}_{i,j}]_{i=0,j=0}^{N_1-1,N_2-1}$ of size $N_1\times N_2$ is said to be a $q$-ary array if $\hat{x}_{i,j} \in \mathbb{Z}_q$ for all $0\leq i\leq N_1-1$ and $0\leq j \leq N_2-1$. Let $\xi_q=e^{-2\pi \sqrt{-1}/q}$ be the $q$-th root of unity. The corresponding complex-valued array  of $\hat{\textit{\textbf{X}}}$ is denoted by $\textit{\textbf{X}}=\left[x_{i,j} \right]_{i=0,j=0}^{N_1-1,N_2-1}$ modulated by $q$-phase-shift keying ($q$-PSK), where $x_{i,j}=\xi_q^{\hat{x}_{i,j}}$.

\subsection{2D $Z$-Complementary Array Quads}
\label{subsec:preliminaries}
Given two complex-valued arrays $\textit{\textbf{X}}$ and $\textit{\textbf{Y}}$ of size $N_1\times N_2$, at 2-D time-shift $(\tau_1,\tau_2)$, the 2-D aperiodic correlation function \cite{2021PaiSPL} of two arrays $\textit{\textbf{X}}$ and $\textit{\textbf{Y}}$ is defined by  
\begin{equation}
	\label{2D-ACCF}
	\rho_{\textbf{\textit{X}}, \textbf{\textit{Y}}}(\tau_1,\tau_2)=
	\sum_{i=0}^{N_1-1-\tau_1}\sum_{j=0}^{N_2-1-\tau_2}x_{i,j} y_{i+\tau_1,j+\tau_2}^{*},
\end{equation}where $0\leq \tau_1< N_1,\ 0\leq \tau_2<N_2$ and $(\cdot)^*$ refers to the complex conjugate. If $\textit{\textbf{X}} =\textit{\textbf{Y}}$, then  $\rho_{\textbf{\textit{X}}, \textbf{\textit{Y}}}(\tau_1,\tau_2)$ is said to be 2-D aperiodic auto-correlation function (2D-AACF); otherwise, it is 2-D aperiodic cross-correlation function (2D-ACCF). The 2D-AACF of $\textbf{\textit{X}}$ is denoted as  $\rho_{\textbf{\textit{X}}}(\tau_1,\tau_2)$. 

When $N_1 = 1$, two 2D arrays  $\textit{\textbf{X}}$ and  $\textit{\textbf{Y}}$ become 1D complex-valued sequences $\textit{\textbf{x}}=(x_{j})_{j=0}^{N_2-1}$ and $\textit{\textbf{y}}=(y_{j})_{j=0}^{N_2-1}$, respectively, with the corresponding 1D-ACCF given by
\begin{equation}
	\rho_{\textbf{\textit{x}},\textbf{\textit{y}}}(\tau)= \sum_{j=0}^{N_2-1-\tau}x_{j} y_{j+\tau}^{*}.
\end{equation} The 1D-AACF of $\textbf{\textit{x}}$ is simply denoted by  $\rho_{\textbf{\textit{x}}}(\tau)$.

\begin{definition}[Golay Complementary Pairs (GCPs)]
	A pair $(\textbf{\textit{x}},\textbf{\textit{y}})$ of 1D sequences with length $N$ is called a Golay complementary pair (GCP) \cite{1999davis} if 
	\begin{align}
		\label{cond:GCP}
		\rho_{\textbf{\textit{x}}}(\tau) +\rho_{\textbf{\textit{y}}}(\tau)=\begin{cases}
			2N, \ \text{if}\ \tau=0 \\
			0, \ \quad \text{if}\ 0< |\tau| \leq N-1.
		\end{cases}
	\end{align}
\end{definition} 
Note that binary GCPs exist only for lengths $2^{\alpha}10^{\beta}26^{\gamma}$ with $\alpha, \beta, \gamma \geq 0$ and $\alpha+\gamma \geq 1-\beta$. However, complex GCPs allow more flexible lengths of the form $2^{a+u}3^b5^c11^d13^e$ with $e,u,a,b,c,d\geq 0, u-c\leq e$ and $(b+d)+(c+e)\leq a+1+2u$  \cite{2002Craigen}.

\begin{definition}[$Z$-Complementary Pairs (ZCPs)]
	A pair $(\textbf{\textit{x}},\textbf{\textit{y}})$ of sequences with length $N$ is called a $Z$-complementary pair (ZCP) \cite{2007Fan_ZCSS} if 
	\begin{align}
		\label{cond:ZCP}
		\rho_{\textbf{\textit{x}}}(\tau) +\rho_{\textbf{\textit{y}}}(\tau)=\begin{cases}
			2N;\ \tau=0 \\
			0; \ \quad  \ 0< |\tau| \leq Z-1,
		\end{cases}
	\end{align}where $Z\leq N$. In this case, $Z$ refers to the ZCZ width. 
\end{definition}Note that a ZCP becomes conventional GCP if $Z=N$. A ZCP with sequence length $N$ and ZCZ width $Z$ will be denoted by $(N,Z)$-ZCP.

\begin{definition}[2D $Z$-Complementary Array Quads]
	A set of $4$ unimodular arrays $\{\textbf{\textit{X}}_1, \textbf{\textit{X}}_2,\textbf{\textit{X}}_3,\textbf{\textit{X}}_4\}$ with size $N_1\times N_2$ is said to be a 2D $Z$-complementary array quad (2D-ZCAQ) if 
	\begin{align}
		\label{2D:ZCAQ:cond:definition}
		&\rho_{\textbf{\textit{X}}_1}(\tau_1,\tau_2) +\rho_{\textbf{\textit{X}}_2}(\tau_1,\tau_2)+\rho_{\textbf{\textit{X}}_3}(\tau_1,\tau_2)+\rho_{\textbf{\textit{X}}_4}(\tau_1,\tau_2) \nonumber \\
		&={\small \begin{cases}
				4N_1N_2; \  \tau_1=\tau_2=0 \\
				0;\  0\leq |\tau_1|< Z_1, 0\leq |\tau_2|< Z_2, (\tau_1,\tau_2)\neq(0,0).
		\end{cases}}
	\end{align}
\end{definition}

	\subsection{Column Sequence PMEPR in MC-CDMA Systems}
For a given $q$-PSK modulated array $\textit{\textbf{X}}=\left[\xi_q^{\hat{x}_{i,j}} \right]_{i=0,j=0}^{N_1-1,N_2-1}$, let $\textit{\textbf{X}}^T_j=\left[X_{0,j},X_{1,j}, \cdots, X_{N_1-1,j} \right]^T$ be the $j$-th column sequence of  $\textit{\textbf{X}}$, where $0\leq j\leq N_2-1$. Then, the time-domain complex baseband OFDM signal can be written by
\begin{align}
	S_{\textbf{\textit{X}}_j^T}(t)
	% &=\sum_{i=0}^{N_1-1}X_{i,j}e^{2\pi \sqrt{-1}(f_c+i\Delta f)t}\nonumber \\
	&=\sum_{i=0}^{N_1-1}\xi_q^{\hat{x}_{i,j}} e^{2\pi \sqrt{-1}(f_c+i\Delta f)t}, \ 0\leq t\leq \frac{1}{\Delta f},
\end{align} where $f_c$ denotes the carrier frequency, $\Delta f$ refers to the subcarrier spacing, and $N_1$ denotes the number of subcarriers in OFDM system.  The instantaneous envelope power ratio (IEPR) of the column sequence $\textbf{\textit{X}}_j^T$ is given by $P_{\textbf{\textit{X}}_j^T}(t)=\frac{\left| S_{\textbf{\textit{X}}_j^T}(t) \right|^2}{N_1}$. Note that $N_1$ is the mean envelope power for $q$-PSK modulated sequences. Then, the peak-to-mean envelope power ratio (PMEPR) \cite{1999davis}, \cite{2000Paterson} of the column sequence $\textbf{\textit{X}}_j^T$ can be defined by
\begin{align}
	\text{PMEPR}\left(\textbf{\textit{X}}_j^T \right)=\text{sup}_{0\leq t\leq \frac{1}{\Delta f}}P_{\textbf{\textit{X}}_j^T}(t).
\end{align}In MC-CDMA system, the column sequence PMEPR value of $\textbf{\textit{X}}_j^T$ should be as low as possible. In this case, the column sequence $\textbf{\textit{X}}_j^T$ is spreaded over the $N_1$ subcarriers in the $j$-th chip-slot \cite{2014ZLiuYGuan}. That is, $X_{i,j}$ is the modulated signal in the $i$-th subcarrier for $i=0,1,\cdots, N_1-1$.

\subsection{Generating Polynomials} 
The generating polynomial \cite{2020DasTSP2D},  \cite{1999FengShyong} of 2D array $\textit{\textbf{X}}$ of size $N_1\times N_2$ over two variables $z_1^{-1}$ and $z_2^{-1}$ is given by
{\small \begin{align}
		\label{array:z:domain}
		X(z_1,z_2)=\sum_{i=0}^{N_1-1}\sum_{j=0}^{N_2-1}x_{i,j} z_1^{-i}z_2^{-j}.
\end{align}} Note that $X(z_1,z_2)$ can also be considered as 2D $z$-transform of the array $\textit{\textbf{X}}$ \cite{2020DasTSP2D}. The 2D polynomial (i.e., $z$-transform) of 2D-AACF $\rho_{\textbf{\textit{X}}}(\tau_1,\tau_2)$ is given by
\begin{align}
	\rho_{\textbf{\textit{X}}}(z_1,z_2)=X(z_1,z_2)\cdot X^*\left(z_1^{-1},z_2^{-1}\right).
\end{align} The polynomial representation of 1D-AACF $\rho_{\textbf{\textit{x}},\textbf{\textit{x}}}(\tau)$ of $\textit{\textbf{x}}$ can be expressed by $ \rho_{\textbf{\textit{x}},\textbf{\textit{x}}}(z)=x(z)\cdot x^*\left(z^{-1}\right)$, where $x(z)$ denotes the polynomial of the 1D sequence $\textbf{\textit{x}}$. The polynomial representation of 1D-AACF $\rho_{\textbf{\textit{x}}}(\tau)$ is simply denoted as  $\rho_{\textbf{\textit{x}}}(z)=x(z)\cdot x^*\left(z^{-1}\right)$. Thus, $\textit{\textbf{x}}$ and $\textit{\textbf{y}}$ form an $(N,Z)$-ZCP if
{\small\begin{align}
	\rho_{\textbf{\textit{x}}}(z)+\rho_{\textbf{\textit{y}}}(z) =2N+\sum_{Z\leq |\tau|<N}(\rho_{\textbf{\textit{x}}}(\tau)+\rho_{\textbf{\textit{y}}}(\tau))z^{-\tau}.
\end{align}}We will use $\tilde{(\cdot)}$ to denote both conjugate and reverse-ordering of a sequence. That is,  $\tilde{\textit{\textbf{x}}}=[x^*(N-1),x^*(N-2),\cdots,x^*(1),x^*(0)].$

\section{New Design Method of Two-Dimensional $Z$-Complementary Array Quads}
\label{sec:prop}
In this section, we first propose a novel construction method for 2D-ZCAQs. Then, we compare the proposed 2D-ZCAQ parameters with some existing works. 

%\subsection{Proposed Construction of 2D-ZCAQs}
\begin{theorem}
	\label{theore:2:2D-ZCAQ}
	Given a 1D-GCP $(x(z), y(z))$ of length $N$ and a 1D $(L,Z)$-ZCP $(a(z),b(z))$, we construct four arrays of equal size $L\times N$ in $z$-domain as follows:
	{\small \begin{align}
		\label{2D:ZCAQ:arrays}
		&X_1(z_1,z_2)=a(z_1)x(z_2); X_2(z_1,z_2)= b(z_1)y(z_2) \nonumber \\
		&X_3(z_1,z_2)=-a(z_1)\tilde{y}(z_2) ;  X_4(z_1,z_2)=b(z_1)\tilde{x}(z_2). 
	\end{align}}Then, the set $\{\textbf{\textit{X}}_1, \textbf{\textit{X}}_2,\textbf{\textit{X}}_3,\textbf{\textit{X}}_4\}$ forms a 2D-ZCAQ with array size $L\times N$ and 2D-ZCZ width $Z\times N$.
\end{theorem}
\begin{IEEEproof}
	According to (\ref{2D:ZCAQ:arrays}), it is clear that each of the arrays is unimodular array of size $L\times N$. Since $(\textbf{\textit{a}},\textbf{\textit{b}})$ is an $(L,Z)$-ZCP and  $(\textbf{\textit{x}},\textbf{\textit{y}})$ is a GCP of length $N$, we have 
	\begin{align}
			&\rho_{\textbf{\textit{x}}}(z)+\rho_{\textbf{\textit{y}}}(z) =2N, \label{GCP:cond:theorem2:Quad}  \\
			&\rho_{\textbf{\textit{a}}}(z)+\rho_{\textbf{\textit{b}}}(z) =2L+\sum_{Z\leq |\tau|<L}(\rho_{\textbf{\textit{a}}}(\tau)+\rho_{\textbf{\textit{b}}}(\tau))z^{-\tau}.\label{ZCP:cond:theorem:Quad}
	\end{align} One can notice that $\rho_{\tilde{\textit{\textbf{y}}}}(z)=\rho_{\textbf{\textit{y}}}(z)$ and $\rho_{\tilde{\textit{\textbf{x}}}}(z)=\rho_{\textbf{\textit{x}}}(z)$. By using (\ref{GCP:cond:theorem2:Quad}) and (\ref{ZCP:cond:theorem:Quad}), we have
	\begin{align}
		&\rho_{\textbf{\textit{X}}_1}(z_1,z_2)+\rho_{\textbf{\textit{X}}_2}(z_1,z_2)+\rho_{\textbf{\textit{X}}_3}(z_1,z_2)+\rho_{\textbf{\textit{X}}_4}(z_1,z_2) \nonumber \\
		%&= (\rho_{\textbf{\textit{a}}}(z_1)+\rho_{\textbf{\textit{b}}}(z_1)) (\rho_{\textbf{\textit{x}}}(z_2)+\rho_{\textbf{\textit{y}}}(z_2))  \nonumber \\
		&=2N(\rho_{\textbf{\textit{a}}}(z_1)+\rho_{\textbf{\textit{b}}}(z_1))  \nonumber \\
		&=4NL+\sum_{Z\leq |\tau_1|<L}2N(\rho_{\textbf{\textit{a}}}(\tau_1)+\rho_{\textbf{\textit{b}}}(\tau_1))z_1^{-\tau_1}. \label{2D:AACF:sum:Quad}
	\end{align} The 2D-ZCZ width is $Z_1\times Z_2$, where $Z_1=Z$ and $Z_2=N$. Thus, the set $\{\textbf{\textit{X}}_1, \textbf{\textit{X}}_2,\textbf{\textit{X}}_3,\textbf{\textit{X}}_4\}$ is a 2D-ZCAQ with array size $L\times N$ and 2D-ZCZ width $Z\times N$. 
\end{IEEEproof} 

Our design method uses a combination of seed sequences formed by 1D-GCPs and 1D-ZCPs. The previous design \cite[Th. 3.5]{2021LiTSP} does not admit such a possibility leading to unimodular 2D-ZCAQs. We compare the set sizes, array sizes and 2D-ZCZ widths for different existing construction methods of 2D-ZCAPs in Table \ref{table:comparison:2D:ZCAQs}.

\begin{remark}
	The number of phases of the constructed 2D-ZCAQs is given by $q=\text{lcm}\{q_0,q_1\}$, where $q_0$ and $q_1$ denote the number of phases for the seed 1D-GCP and 1D-ZCP, respectively.
\end{remark}

\begin{example}
	\label{example:2d:zcaq}
	Let $N=3$ and $L=7$ with $Z=4$. We consider a length-$3$ four-phase GCP $(\textbf{\textit{x}}, \textbf{\textit{y}})$ with $\textbf{\textit{x}}=[+ \ +\ -]$ and $\textbf{\textit{y}}=[+\ j\ +]$, and a binary $(7,4)$-ZCP $(\textbf{\textit{a}}, \textbf{\textit{b}})$ with  $\textbf{\textit{a}}=[+ + + + - - +]$ and $\textbf{\textit{b}}=[+ + - + - + +]$, where $+$ and $-$ represent $+1$ and $-1$, respectively. We have four $7\times 3$ arrays $\textbf{\textit{X}}_1, \textbf{\textit{X}}_2,\textbf{\textit{X}}_3$ and $\textbf{\textit{X}}_4$ given by \newline {\footnotesize $\textbf{\textit{X}}_1^T=\begin{bmatrix}
			++++--+\\
			++++--+\\
			----++-
		\end{bmatrix},$ $ \textbf{\textit{X}}_2^T=\begin{bmatrix}
				+\ \ +\ \ -\ \ +\ \ - \ \ + \ \ +\\
				j \ \ \ j \  -j \ \ \ j \  -j \ \ \ j \ \ \ j\\
				+\ \ +\ \ -\ \ + \ \ - \ \ + \ \ +
			\end{bmatrix}$, $\textbf{\textit{X}}_3^T=\begin{bmatrix}
			-\ \ -\ \ -\ \ -\ \ + \ \ + \ \ -\\
			j \,\ \ j \ \ \ j \ \ \ j  \ -j  \ -j \ \ \ j\\
			-\ \ -\ \ -\ \ - \ \ + \ \ + \ \ - 
		\end{bmatrix}$, $\textbf{\textit{X}}_4^T=\begin{bmatrix}
				--+-+--\\
				++-+-++\\
				++-+-++
			\end{bmatrix}$}. In Fig. \ref{fig:2D-AACF:sum:2d:zcaq}, we show the absolute values of 2D-AACF sums for four arrays $\textbf{\textit{X}}_1,$ $ \textbf{\textit{X}}_2,$ $\textbf{\textit{X}}_3$ and $\textbf{\textit{X}}_4$. The number of phases of the constructed array is $q=\text{lcm}\{2,4\}=4$. Thus, the set $\{\textbf{\textit{X}}_1, \textbf{\textit{X}}_2,\textbf{\textit{X}}_3,\textbf{\textit{X}}_4\}$ forms a $4$-ary 2D-ZCAQ with array size $7\times 3$ and 2D-ZCZ width $4\times 3$.
	\begin{center}
		\begin{figure}
			\includegraphics[width=0.8\textwidth]{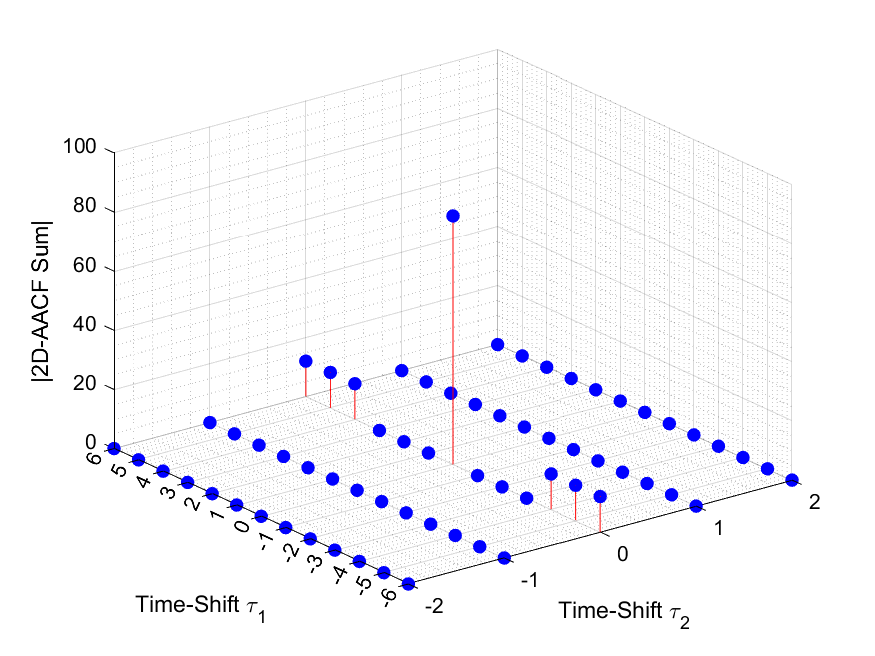}
			\caption{The sum of 2D-AACFs from \textit{Example \ref{example:2d:zcaq}}}
			\label{fig:2D-AACF:sum:2d:zcaq}
		\end{figure}
	\end{center}
\end{example}

\begin{figure*}
	\begin{center}
		\begin{tabular}{l}
			\captionof{table}{A Summary of Existing Works}  \label{table:comparison:2D:ZCAQs}
		\end{tabular}
	\end{center}
\end{figure*}
\begin{figure*}
	\begin{center}
		\resizebox{\linewidth}{!}{
			\renewcommand{\arraystretch}{1.11}
			\begin{tabular}{|l|l|l|l|l|l|l|l|l|l|l|}
				\hline
				\multicolumn{1}{|c||}{Refs.}  & \multicolumn{1}{|c||}{Phases}&\multicolumn{1}{c||}{Set Size}& \multicolumn{1}{c||}{Array Size} & \multicolumn{1}{c||}{2D-ZCZ Width} & \multicolumn{1}{c||}{Constraints} & \multicolumn{1}{c|}{Based On} \\ 
				\hline
				\hline
				\multicolumn{1}{|c||}{\cite{2021PaiTIT}} & \multicolumn{1}{|c||}{$2$}  &\multicolumn{1}{c||}{$2$}&  \multicolumn{1}{c||}{$L_1\times L_2$} & \multicolumn{1}{c||}{$Z_1\times Z_2$} & \multicolumn{1}{c||}{$L_1,L_2\geq 2$} & \multicolumn{1}{c|}{$(L_1,Z_1)$-ZCP and $(L_2,Z_2)$-ZCP} \\ 
				\hline
				\multicolumn{1}{|c||}{\cite{2021AbhishekCOML}} & \multicolumn{1}{|c||}{$q$}  &\multicolumn{1}{c||}{$2$}& \multicolumn{1}{c||}{$(2^{m_1-1}+2^{n+1})\times (2^{m_2}+4)$} & \multicolumn{1}{c||}{$(2^{\pi(n+1)}+2^{n+1})\times (2^{m_2-2}+2^{\phi(m_2-3)}+1)$} & \multicolumn{1}{c||}{$2|q;m_1\geq 2;\ n\leq m_1-3;\ m_2\geq 4$} & \multicolumn{1}{c|}{Generalized Boolean Functions} \\ 
				\hline
				\multicolumn{1}{|c||}{\cite{2021PaiSPL}} & \multicolumn{1}{|c||}{$q$}  &\multicolumn{1}{c||}{$2$}&  \multicolumn{1}{c||}{$2^n\times (2^{m-1}+\sum_{\alpha=t+1}^{m-1}d_\alpha 2^{\alpha-1}+2^\nu)$} & \multicolumn{1}{c||}{$2^n\times (2^{t-1}+2^\nu)$} & \multicolumn{1}{c||}{$2|q;m,n,t,\nu \geq 0; \nu<t<m; d_\alpha \in \{0,1\}$} & \multicolumn{1}{c|}{2D Generalized Boolean Functions} \\ 
				\hline
				\multicolumn{1}{|c||}{\cite[Th. 2]{2021Zhang}} & \multicolumn{1}{|c||}{$q$}  &\multicolumn{1}{c||}{$2$}& \multicolumn{1}{c||}{$14\cdot 2^n\times 2^{m-n}$} & \multicolumn{1}{c||}{$12\cdot 2^n\times 2^{m-n}$} & \multicolumn{1}{c||}{$2|q;0\leq n\leq m$} & \multicolumn{1}{c|}{2D Generalized Boolean Functions} \\ 
				\cline{1-7}
				\multicolumn{1}{|c||}{\cite[Lem. 5]{2021Zhang}} & \multicolumn{1}{|c||}{$q$}  &\multicolumn{1}{c||}{$2$}& \multicolumn{1}{c||}{$2^n\times (2^{m-1}+\sum_{\alpha=t+1}^{m-1}d_\alpha 2^{\alpha-1}+2^\nu)$} & \multicolumn{1}{c||}{$2^n\times (2^{t-1}+2^\nu)$} & \multicolumn{1}{c||}{$2|q;m,n,t,\nu \geq 0; \nu<t<m; d_\alpha \in \{0,1\}$} & \multicolumn{1}{c|}{Generalized Boolean Functions} \\ 
				\hline
				\multicolumn{1}{|c||}{Proposed} & \multicolumn{1}{|c||}{$q$}  & \multicolumn{1}{c||}{$4$}& \multicolumn{1}{c||}{$L\times 2^{\alpha}10^{\beta}26^{\gamma}$} & \multicolumn{1}{c||}{$Z\times 2^{\alpha}10^{\beta}26^{\gamma}$} & \multicolumn{1}{c||}{$q=\text{lcm}\{2,q_1\}, q_1\geq 2$} & \multicolumn{1}{c|}{Binary GCP and $(L,Z)$-ZCP} \\ 
				\cline{2-7}
				\multicolumn{1}{|c||}{} & \multicolumn{1}{|c||}{$q$}  & \multicolumn{1}{c||}{$4$}&  \multicolumn{1}{c||}{$L\times 2^{a+u}3^b5^c11^d13^e$} & \multicolumn{1}{c||}{$Z\times 2^{a+u}3^b5^c11^d13^e$} & \multicolumn{1}{c||}{$q=\text{lcm}\{q_0,q_1\}, q_0,q_1\geq 2$} & \multicolumn{1}{c|}{Complex GCP and $(L,Z)$-ZCP} \\ 
				\hline
		\end{tabular}}
	\end{center}
\end{figure*}

\section{The Column Sequence PMEPR Values of 2D $Z$-Complementary Array Quads}
\label{sec:PMEPR:2D-ZCAQs}
For any sequence pair  $(\textbf{\textit{p}}, \textbf{\textit{q}})$ of length $L$, there is a well-known connection between the auto-correlation properties of $\textbf{\textit{p}}$ and $\textbf{\textit{q}}$ and their PMEPR values given by \cite{2014Liu_TIT}
\begin{align}
	\label{PMEPR:any:pair}
	\text{PMEPR}(\textbf{\textit{p}})\leq 2+\frac{2}{L} \sum_{\tau=1}^{L-1}\Big| \rho_{\textbf{\textit{p}}}(\tau)+	\rho_{\textbf{\textit{q}}}(\tau)   \Big|.
\end{align} In what follows, we provide a theorem to show that the column sequence PMEPR values of the constructed 2D-ZCAQs from \textit{Theorem \ref{theore:2:2D-ZCAQ}} is connected only to the specific auto-correlation properties of the seed $(L,Z)$-ZCP $(\textbf{\textit{a}},\textbf{\textit{b}})$.

\begin{theorem}
	\label{th:PMEPR}
The column sequence PMEPR values of the constructed 2D-ZCAQs from \textit{Theorem \ref{theore:2:2D-ZCAQ}}  are given by 
\begin{align}
	\label{PMEPR:Xm:th}
	\text{PMEPR}(X_{m,j}^T)\leq 2+\frac{2}{L} \sum_{\tau=1}^{L-1}\Big|  \rho_{\textbf{\textit{a}}}(\tau)+	\rho_{\textbf{\textit{b}}}(\tau)  \Big|, 
\end{align} where $X_{m,j}^T$ is the $j$-th column sequence of the array $\textbf{\textit{X}}_m$ with $m=1,2,3,4$ and $j=0,1,\cdots, N-1$.
\end{theorem}

\begin{IEEEproof}
According to (\ref{2D:ZCAQ:arrays}), the array  $\textbf{\textit{X}}_1$ of size $L\times N$ can be written in time-domain as follows:
\begin{align}
	\textbf{\textit{X}}_1
	%&=\begin{bmatrix}
	%	a_0x_0 & a_0x_1 & \cdots & a_0x_{N-1} \\
	%	a_1x_0 & a_1x_1 & \cdots & a_1x_{N-1} \\
	%	\vdots & \vdots & \ddots & \vdots \\
	%	a_{L-1}x_0 & a_{L-1}x_1 & \cdots & a_{L-1}x_{N-1} \\
%	\end{bmatrix}  \nonumber \\
&=\begin{bmatrix}
	x_0\textbf{\textit{a}}^T & x_1\textbf{\textit{a}}^T & \cdots & x_{N-1} \textbf{\textit{a}}^T 
\end{bmatrix} \nonumber \\
&=[X_{1,0},X_{1,1},\cdots, X_{1,N-1}]
\end{align}Therefore, the $j$-th column sequence $X_{1,j}^T$ of the array $\textbf{\textit{X}}_1$ is given by $X_{1,j}^T=x_j\textbf{\textit{a}}$ for $j=0,1,\cdots, N-1$. Similarly, the $j$-th column sequence $X_{2,j}^T$, $X_{3,j}^T$ and $X_{4,j}^T$ of the array $\textbf{\textit{X}}_2$, $\textbf{\textit{X}}_3$, and $\textbf{\textit{X}}_4$, respectively, can be written by
\begin{align}
	X_{2,j}^T=y_j\textbf{\textit{b}}; \ X_{3,j}^T=-y^*_{N-1-j}\textbf{\textit{a}}\ \text{and}\ X_{4,j}^T=x^*_{N-1-j}\textbf{\textit{b}}. 
\end{align} According to (\ref{PMEPR:any:pair}), the column sequence PMEPR value for the sequence pair $(X_{1,j}^T, X_{4,N-1-j}^T)$ of length $L$ is given by
\begin{align}
	\label{PMEPR:X1}
		\text{PMEPR}(X_{1,j}^T)
	%	&\leq 2+\frac{2}{L} \sum_{\tau=1}^{L-1}\Big| \rho_{X_{1,j}^T}(\tau)+	\rho_{X_{4,N-1-j}^T}(\tau)   \Big|\nonumber \\
		&\leq 2+\frac{2}{L} \sum_{\tau=1}^{L-1}\Big|  x_j^2(\rho_{\textbf{\textit{a}}}(\tau)+	\rho_{\textbf{\textit{b}}}(\tau))   \Big|\nonumber \\
		&\leq 2+\frac{2}{L} \sum_{\tau=1}^{L-1}\Big|  \rho_{\textbf{\textit{a}}}(\tau)+	\rho_{\textbf{\textit{b}}}(\tau)  \Big|, 
\end{align} where $|x_j|=1$. Similarly, the column sequence PMEPR value for the sequence pair $(X_{2,j}^T, X_{3,N-1-j}^T)$ of length $L$ is given by
\begin{align}
	\label{PMEPR:X2}
	\text{PMEPR}(X_{2,j}^T)\leq 2+\frac{2}{L} \sum_{\tau=1}^{L-1}\Big|  \rho_{\textbf{\textit{a}}}(\tau)+	\rho_{\textbf{\textit{b}}}(\tau)  \Big|, 
\end{align}where $|y_j|=1$. This completes the proof. 
\end{IEEEproof}
\begin{remark}
	Based on \textit{Theorem \ref{th:PMEPR}}, the PMEPR values are calculated by using only unimodular seed 1D-GCPs and 1D-ZCPs in  \textit{Theorem \ref{theore:2:2D-ZCAQ}}. 
\end{remark}

\begin{proposition}
	\label{prop:pmepr:3.33}
	The column sequence PMEPR of the constructed 2D-ZCAQs by \textit{Theorem \ref{theore:2:2D-ZCAQ}}  is upper bounded by $3.33$ when a seed $(L,Z)$-ZCP from \cite{2014Liu_Z_CCS} is used with $L=2^{n+1}+2^{n}$ and $Z=2^{n+1}$.
\end{proposition}
\begin{IEEEproof}
	In  \cite[Th. 3]{2014Liu_Z_CCS}, it has been shown that the constructed $(L,Z)$-ZCP $(\textbf{\textit{a}},\textbf{\textit{b}})$ from \cite{2014Liu_Z_CCS}  has the following auto-correlation property given by
	 \begin{align}
	 	&\rho_{\textbf{\textit{a}}}(\tau)+	\rho_{\textbf{\textit{b}}}(\tau) =\begin{cases}
	 		2^{n+2}+2^{n+1} , \ \text{if} \ \tau=0,  \\
	 		\pm 2^{n+1},\ \text{if}\ \tau=2^{n+1}, \\
	 		0, \ \text{otherwise}.
	 	\end{cases}
	 \end{align} By using (\ref{PMEPR:X1}), the $j$-th column sequence $X_{1,j}^T$ of the array $\textbf{\textit{X}}_1$ is given by 
 \begin{align}
 	\label{upprbound:3.33}
 \text{PMEPR}(X_{1,j}^T)\leq 2+\frac{2}{2^{n+1}+2^{n}}\cdot 2^{n+1} 
 =2+\frac{4}{3} \approx 3.33
\end{align} for $j=0,1,\cdots, N-1$. Similarly, we can show that $\text{PMEPR}(X_{2,j}^T)\leq 3.33$, $\text{PMEPR}(X_{3,N-1-j}^T)\leq 3.33$ and $\text{PMEPR}(X_{4,N-1-j}^T)\leq 3.33$. This completes the proof.
\end{IEEEproof}
\begin{center}
	\begin{figure}
		\includegraphics[width=0.75\textwidth]{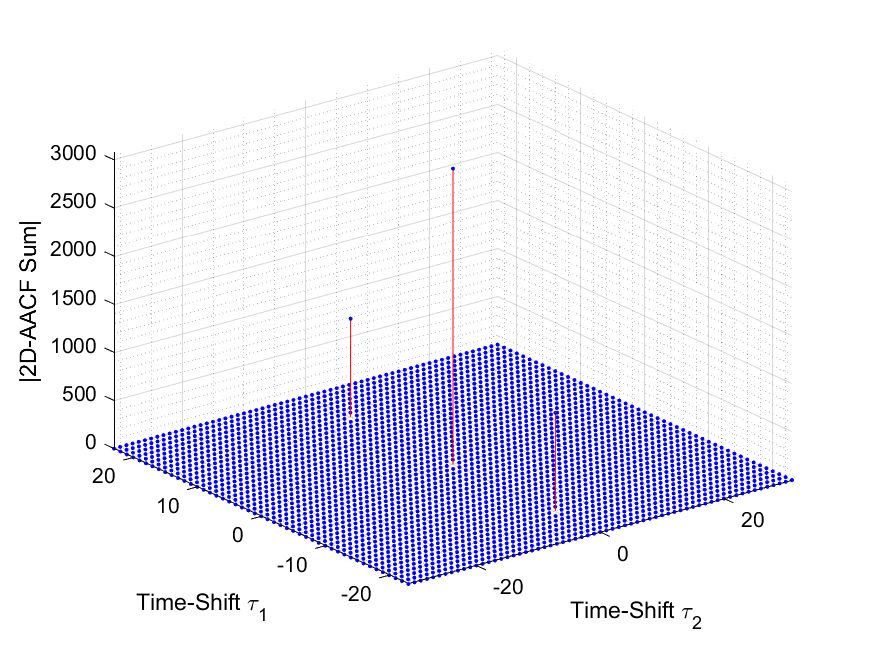}
		\caption{The sum of 2D-AACFs from \textit{Example \ref{example2:2d:zcaq}}}
		\label{fig:2D-AACF:sum:2d:zcaq:exxam2}
	\end{figure}
\end{center}
\begin{center}
	\begin{figure}
		\includegraphics[width=0.75\textwidth]{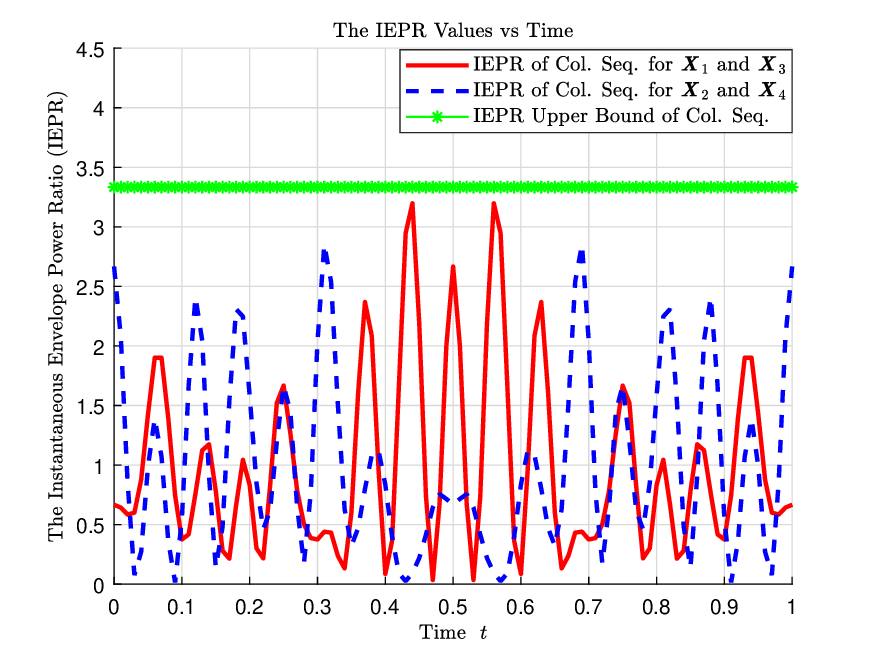}
		\caption{The IEPRs of column sequences for $\textbf{\textit{X}}_1, \textbf{\textit{X}}_2, \textbf{\textit{X}}_3$ and $\textbf{\textit{X}}_4$  given in \textit{Example \ref{example2:2d:zcaq}}}
		\label{fig:2D-AACF:colum:pmepr:exxam2}
	\end{figure}
\end{center}
\begin{example}
	\label{example2:2d:zcaq}
	Let $n=3$ with $L=24, Z=16$ and $N=32$. Let us consider a $(24,16)$-ZCP $(\textbf{\textit{a}},\textbf{\textit{b}})$ constructed  from \cite{2014Liu_Z_CCS} with $\textbf{\textit{a}}=[+-+-++-------++-+-+-++--]$ and $\textbf{\textit{b}}=[-++-----++--+-+--++-----]$. We take a GCP $(\textbf{\textit{x}},\textbf{\textit{y}})$ of length $32$ with $\textbf{\textit{x}}=[+ ++++ -+ - -    -+++ --+++-  -     +    -    -     +    -    -    -    -     +   -    +    -]$ and $\textbf{\textit{y}}=[-    -    -    -    -     +  -     +    +     +    -    -    -     +     +    -     +     +    -    -     +    -    -     +    -    -    -    -     +    -      +  -]$. Based on \textit{Theorem \ref{theore:2:2D-ZCAQ}}, we construct four arrays $\textbf{\textit{X}}_1, \textbf{\textit{X}}_2,\textbf{\textit{X}}_3$ and $\textbf{\textit{X}}_4$ of equal size $24\times 32$. We show the absolute values of 2D-AACF sums for four arrays $\textbf{\textit{X}}_1,$ $ \textbf{\textit{X}}_2,$ $\textbf{\textit{X}}_3$ and $\textbf{\textit{X}}_4$ in Fig. \ref{fig:2D-AACF:sum:2d:zcaq:exxam2}. We can see that the set $\{\textbf{\textit{X}}_1, \textbf{\textit{X}}_2,\textbf{\textit{X}}_3,\textbf{\textit{X}}_4\}$ forms a binary 2D-ZCAQ with array size $24\times 32$ and 2D-ZCZ width $16\times 32$. The maximum column sequence PMEPR of each column of $\textbf{\textit{X}}_1$ and $\textbf{\textit{X}}_3$ is 3.197. The maximum column sequence PMEPR of each column of $\textbf{\textit{X}}_2$ and $\textbf{\textit{X}}_4$ is 2.851. For example, we consider the column sequence IEPR of the first column of $\textbf{\textit{X}}_2$ as illustrated in Fig. \ref{fig:2D-AACF:colum:pmepr:exxam2}. Note that the sub-carrier spacing is normalized to $1$ for Fig. \ref{fig:2D-AACF:colum:pmepr:exxam2}. According to Fig. \ref{fig:2D-AACF:colum:pmepr:exxam2}, we have 
	\begin{align}
		\text{PMEPR}(X_{2,0}^T)=\text{sup}_{0\leq t\leq 1}\text{IEPR}_{X_{2,0}^T}(t)=2.851,
	\end{align} whereas the theoretical PMEPR upper bound is 3.33 (given by (\ref{upprbound:3.33})) from \textit{Proposition \ref{prop:pmepr:3.33}}. Similarly, the maximum column sequence PMEPR of each column of $\textbf{\textit{X}}_1$ is 3.197 as observed in Fig. \ref{fig:2D-AACF:colum:pmepr:exxam2}. 
\end{example}
\begin{proposition}
	\label{prop:pmepr:4}
	The column sequence PMEPR of the constructed 2D-ZCAQs by \textit{Theorem \ref{theore:2:2D-ZCAQ}} is upper bounded by $4$ when a seed $(L,Z)$-ZCP from \cite{2018AvikSPL} is used with $L=2N+2$ and $Z=3N/2+1$.
\end{proposition}

\begin{IEEEproof}
	By using \cite[(16), (17) \text{and} (18)]{2018AvikSPL}, the constructed $(L,Z)$-ZCP $(\textbf{\textit{a}},\textbf{\textit{b}})$ from \cite{2018AvikSPL} has the following auto-correlation property given by
	\begin{align}
		&\rho_{\textbf{\textit{a}}}(\tau)+	\rho_{\textbf{\textit{b}}}(\tau) =\begin{cases}
		%	4N+4 , \ \text{when} \ \tau=0,  \\
			0, \ \text{if} \ 0<\tau \leq 3N/2 \\
			\pm 4,\ \text{if}\ 3N/2<\tau\leq 2N, \\
			0,  \ \text{if} \ \tau=2N+1, 
		\end{cases}
	\end{align} Based on (\ref{PMEPR:X1}), the $j$-th column sequence $X_{1,j}^T$ of the array $\textbf{\textit{X}}_1$ is given by 
\begin{align}
\text{PMEPR}(X_{1,j}^T)&\leq 2+\frac{2}{2N+2}\sum_{\tau=3N/2+1}^{2N}4 <4
\end{align} for $j=0,1,\cdots, N-1$. Similarly, we can show that $\text{PMEPR}(X_{2,j}^T)\leq 4$, $\text{PMEPR}(X_{3,N-1-j}^T)\leq 4$ and $\text{PMEPR}(X_{4,N-1-j}^T)\leq 4$. This completes the proof.
\end{IEEEproof}

\begin{center}
	\begin{figure}
		\includegraphics[width=0.8\textwidth]{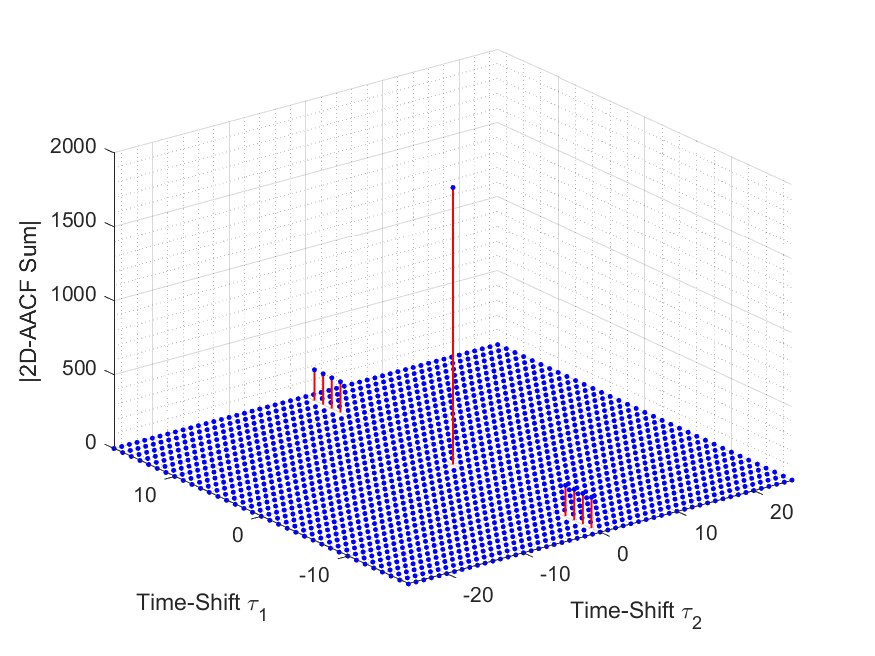}
		\caption{The sum of 2D-AACFs from \textit{Example \ref{example3:2d:pmepr}}}
		\label{fig:2D-AACF:sum:2d:exxam3}
	\end{figure}
\end{center}

\begin{center}
	\begin{figure}
		\includegraphics[width=0.75\textwidth]{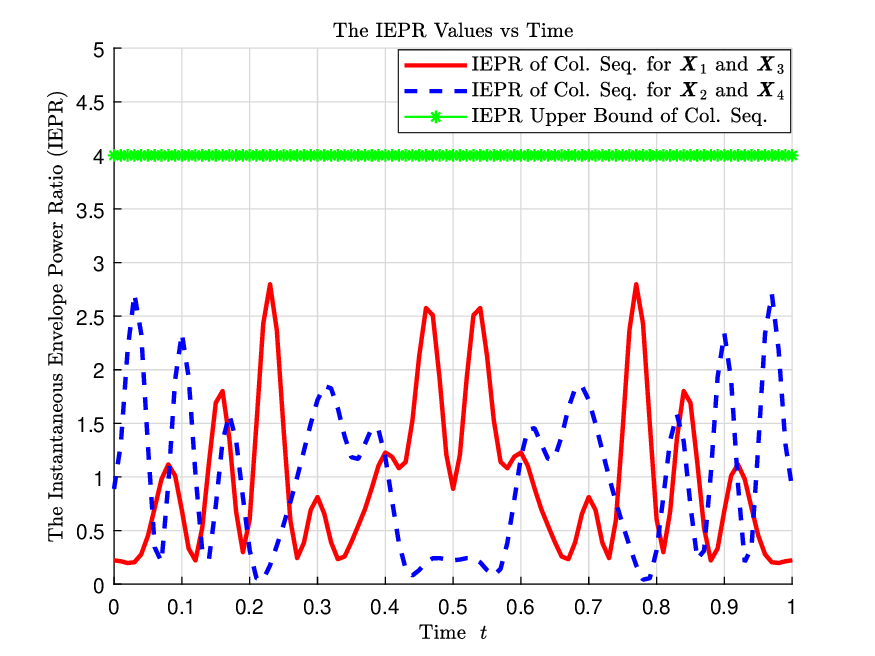}
		\caption{The IEPRs of column sequences for $\textbf{\textit{X}}_1, \textbf{\textit{X}}_2, \textbf{\textit{X}}_3$ and $\textbf{\textit{X}}_4$ given in \textit{Example \ref{example3:2d:pmepr}}}
		\label{fig:2D-AACF:pmepr:exxam3}
	\end{figure}
\end{center}

\begin{example}
	\label{example3:2d:pmepr}
	Let $L=18, Z=13$ and $N=26$. Let us take a binary $(18,13)$-ZCP $(\textbf{\textit{a}},\textbf{\textit{b}})$ constructed  from \cite{2018AvikSPL} with $\textbf{\textit{a}}=[-+++ - + + - + - + - - -  +++ -]$ and $\textbf{\textit{b}}=[++++ - - - + - - + - - + - - - -]$. We consider a binary GCP $(\textbf{\textit{x}},\textbf{\textit{y}})$ of length $26$ with $\textbf{\textit{x}}=[++++ - + + - - + - + - + - - + - + + + - - + + +]$ and $\textbf{\textit{y}}=[++++ - + + - - + - + + + + + - + - - - + + - - -]$. According to \textit{Theorem \ref{theore:2:2D-ZCAQ}}, we construct four binary arrays $\textbf{\textit{X}}_1, \textbf{\textit{X}}_2,\textbf{\textit{X}}_3$ and $\textbf{\textit{X}}_4$ of equal size $18\times 26$. The absolute values of 2D-AACF sums for four arrays $\textbf{\textit{X}}_1,$ $ \textbf{\textit{X}}_2,$ $\textbf{\textit{X}}_3$ and $\textbf{\textit{X}}_4$ are shown in Fig. \ref{fig:2D-AACF:sum:2d:exxam3}. We can observe that the set $\{\textbf{\textit{X}}_1, \textbf{\textit{X}}_2,\textbf{\textit{X}}_3,\textbf{\textit{X}}_4\}$ forms a binary 2D-ZCAQ with array size $18\times 26$ and 2D-ZCZ width $13\times 26$.	The maximum column sequence PMEPR of each column of $\textbf{\textit{X}}_1$ and $\textbf{\textit{X}}_3$ is 2.797. The maximum column sequence PMEPR of each column of $\textbf{\textit{X}}_2$ and $\textbf{\textit{X}}_4$ is 2.706. For example, we consider the column sequence IEPR of the first column of $\textbf{\textit{X}}_1$ as illustrated in Fig. \ref{fig:2D-AACF:pmepr:exxam3}.  Based on Fig. \ref{fig:2D-AACF:pmepr:exxam3}, we have 
	\begin{align}
		\text{PMEPR}(X_{1,0}^T)=\text{sup}_{0\leq t\leq 1}\text{IEPR}_{X_{1,0}^T}(t)=2.797,
	\end{align} whereas the theoretical PMEPR upper bound is 4 from \textit{Proposition \ref{prop:pmepr:4}}. Similarly, the maximum column sequence PMEPR of each column of $\textbf{\textit{X}}_2$ is 2.706 as depicted in Fig. \ref{fig:2D-AACF:pmepr:exxam3}. 
\end{example}

\begin{proposition}
	The column sequence PMEPR of the constructed 2D-ZCAQs by \textit{Theorem \ref{theore:2:2D-ZCAQ}} is upper bounded by $3.714$ when a seed $(L,Z)$-ZCP from \cite[Th. 1]{2018Xie} is used with $L=2^{n+3}+2^{n+2}+2^{n+1}$ and $Z=2^{n+3}$.
\end{proposition}
\begin{IEEEproof}
	The constructed $(L,Z)$-ZCP $(\textbf{\textit{a}},\textbf{\textit{b}})$ from \cite[Th. 1]{2018Xie} has the following auto-correlation property \cite{2019Chen} given by 
	\begin{align}
		&\rho_{\textbf{\textit{a}}}(\tau)+	\rho_{\textbf{\textit{b}}}(\tau) \nonumber \\
		&=\begin{cases}
			2^{n+4}+2^{n+3}+2^{n+2} , \ \text{when} \ \tau=0,  \\
			\pm 2^{n+2},\ \text{when}\ \tau=2^{n+3}+l\cdot 2^{n+1} \ \text{for}\ l=0,1,2, \\
			0, \ \text{otherwise}.
		\end{cases}
	\end{align}According to (\ref{PMEPR:X1}), the $j$-th column sequence $X_{1,j}^T$ of the array $\textbf{\textit{X}}_1$ is given by 
	\begin{align}
		\text{PMEPR}(X_{1,j}^T)&\leq 2+\frac{2}{2^{n+3}+2^{n+2}+2^{n+1}}\cdot 3\cdot 2^{n+2} \nonumber \\
		&=2+\frac{12}{7}\approx 3.714
	\end{align} for $j=0,1,\cdots, N-1$. Similarly, we can show that $\text{PMEPR}(X_{2,j}^T)\leq 3.714$, $\text{PMEPR}(X_{3,N-1-j}^T)\leq 3.714$ and $\text{PMEPR}(X_{4,N-1-j}^T)\leq 3.714$. 
\end{IEEEproof}


\section{Conclusion}
\label{sec:conclusion}
In this paper, we have investigated a new construction for 2D-ZCAQs with various array sizes $L\times N$, where $N=2^{\alpha}10^{\beta}26^{\gamma}$ or $2^{a+u}3^b5^c11^d13^e$ and $L$ is any positive integer $\geq 2$. The admissible lengths of 1D-GCPs and 1D-ZCPs allow us to offer more feasible array sizes of 2D-ZCAQs. Furthermore, we have shown that the column sequence PMEPR of the proposed 2D-ZCAQ is upper bounded by $4$ when we use seed 1D-ZCPs from \cite{2014Liu_Z_CCS}, \cite{2018AvikSPL} and \cite[Th. 1]{2018Xie}. The proposed 2D-ZCAQs with low column sequence PMEPRs can be utilized as alternatives to 2D-GCASs in UWB MC-CDMA systems.

%%Bibliography
%\bibliographystyle{IEEEtran}
%\bibliography{refers}

\begin{thebibliography}{10}
	\providecommand{\url}[1]{#1}
	\csname url@samestyle\endcsname
	\providecommand{\newblock}{\relax}
	\providecommand{\bibinfo}[2]{#2}
	\providecommand{\BIBentrySTDinterwordspacing}{\spaceskip=0pt\relax}
	\providecommand{\BIBentryALTinterwordstretchfactor}{4}
	\providecommand{\BIBentryALTinterwordspacing}{\spaceskip=\fontdimen2\font plus
		\BIBentryALTinterwordstretchfactor\fontdimen3\font minus
		\fontdimen4\font\relax}
	\providecommand{\BIBforeignlanguage}[2]{{%
			\expandafter\ifx\csname l@#1\endcsname\relax
			\typeout{** WARNING: IEEEtran.bst: No hyphenation pattern has been}%
			\typeout{** loaded for the language `#1'. Using the pattern for}%
			\typeout{** the default language instead.}%
			\else
			\language=\csname l@#1\endcsname
			\fi
			#2}}
	\providecommand{\BIBdecl}{\relax}
	\BIBdecl
	
	\bibitem{1961golay}
	M.~J.~E. Golay, ``Complementary {S}eries,'' \emph{IRE Trans. Inf. Theory}, vol.
	IT-7, no.~2, pp. 82--87, Apr. 1961.
	
	\bibitem{1999davis}
	J.~A. Davis and J.~Jedwab, ``Peak-to-mean power control in {OFDM}, {G}olay
	complementary sequences, and {R}eed-{M}uller codes,'' \emph{IEEE Trans. Inf.
		Theory}, vol.~45, no.~7, pp. 2397--2417, Nov. 1999.
	
	\bibitem{2000Paterson}
	K.~G. Paterson, ``Generalized {R}eed-{M}uller codes and power control in {OFDM}
	modulation,'' \emph{IEEE Trans. Inf. Theory}, vol.~46, no.~1, pp. 104--120,
	2000.
	
	\bibitem{2001Spasojevic}
	P.~Spasojevic and C.~N. Georghiades, ``Complementary sequences for {ISI}
	channel estimation,'' \emph{IEEE Trans. Inf. Theory}, vol.~47, no.~3, pp.
	1145--1152, 2001.
	
	\bibitem{1990Budishin}
	S.~Z. Budi\v{s}in, ``New complementary pairs of sequences,'' \emph{Electron.
		Lett.}, vol.~26, pp. 881--883(2), June 1990.
	
	\bibitem{2007Fan_ZCSS}
	P.~Fan, W.~Yuan, and Y.~Tu, ``Z-complementary binary sequences,'' \emph{IEEE
		Signal Process. Lett.}, vol.~14, no.~8, pp. 509--512, Aug. 2007.
	
	\bibitem{2011LiExiztenceZCP}
	X.~Li, P.~Fan, X.~Tang, and Y.~Tu, ``Existence of binary {Z}-complementary
	pairs,'' \emph{IEEE Signal Process. Lett.}, vol.~18, no.~1, pp. 63--66, 2011.
	
	\bibitem{2014Liu_TIT}
	Z.~Liu, U.~Parampalli, and Y.~L. Guan, ``Optimal odd-length binary
	{Z}-complementary pairs,'' \emph{IEEE Trans. Inf. Theory}, vol.~60, no.~9,
	pp. 5768--5781, Sep. 2014.
	
	\bibitem{2014Liu_Z_CCS}
	------, ``On even-period binary {Z}-complementary pairs with large {ZCZ}s,''
	\emph{IEEE Signal Process. Lett.}, vol.~21, no.~3, pp. 284--287, Mar. 2014.
	
	\bibitem{2017Chen}
	C.~Chen, ``A novel construction of {Z}-complementary pairs based on generalized
	{B}oolean functions,'' \emph{IEEE Signal Process. Lett.}, vol.~24, no.~7, pp.
	987--990, Jul. 2017.
	
	\bibitem{2017APCCShibsankar}
	S.~Das, S.~Majhi, S.~Budi\v{s}in, Z.~Liu, and Y.~L. Guan, ``A novel
	multiplier-free generator for complete complementary codes,'' in \emph{Proc.
		23rd Asia-Pacific Conference on Communications (APCC)}, Dec. 2017.
	
	\bibitem{2017Shibsankar}
	S.~Das, S.~Budi\v{s}in, S.~Majhi, Z.~Liu, and Y.~L. Guan, ``A multiplier-free
	generator for polyphase complete complementary codes,'' \emph{IEEE Trans.
		Signal Process.}, vol.~66, no.~5, pp. 1184--1196, Nov. 2017.
	
	\bibitem{2018Xie}
	C.~Xie and Y.~Sun, ``Constructions of even-period binary {Z}-complementary
	pairs with large {ZCZ}s,'' \emph{IEEE Signal Process. Lett.}, vol.~25, no.~8,
	pp. 1141--1145, Aug. 2018.
	
	\bibitem{2018AvikSPL}
	A.~R. Adhikary, S.~Majhi, Z.~Liu, and Y.~L. Guan, ``New sets of even-length
	binary {Z}-complementary pairs with asymptotic {ZCZ} ratio of $3/4$,''
	\emph{IEEE Signal Process. Lett.}, vol.~25, no.~7, pp. 970--973, Jul. 2018.
	
	\bibitem{2018Shibsankar}
	S.~Das, S.~Majhi, and Z.~Liu, ``A novel class of complete complementary codes
	and their applications for {APU} matrices,'' \emph{IEEE Signal Process.
		Lett.}, vol.~25, no.~9, pp. 1300--1304, Jul. 2018.
	
	\bibitem{2019ShibsankarTSP}
	S.~Das, S.~Majhi, S.~Budi\v{s}in, and Z.~Liu, ``A new construction framework
	for polyphase complete complementary codes with various lengths,'' \emph{IEEE
		Trans. Signal Process.}, vol.~67, no.~10, pp. 2639--2648, Mar. 2019.
	
	\bibitem{2019Chen}
	C.~Chen and C.~Pai, ``Binary {Z}-complementary pairs with bounded peak-to-mean
	envelope power ratios,'' \emph{IEEE Commun. Lett.}, vol.~23, no.~11, pp.
	1899--1903, Nov. 2019.
	
	\bibitem{2019ShibIWSDA}
	S.~Das, U.~Parampalli, S.~Majhi, and Z.~Liu, ``Construction of new optimal
	{Z}-complementary code sets from {Z}-paraunitary matrices,'' in \emph{IEEE
		International Workshop on Signal Design and its Applications in
		Communications (IWSDA)}, 2019, pp. 1--5.
	
	\bibitem{2020Shibsankar_TSP2020_II}
	S.~{Das}, U.~{Parampalli}, S.~{Majhi}, Z.~{Liu}, and S.~{Budišin}, ``New
	optimal {$Z$}-complementary code sets based on generalized paraunitary
	matrices,'' \emph{IEEE Trans. Signal Process.}, vol.~68, pp. 5546--5558,
	2020.
	
	\bibitem{2022ISITShib}
	S.~Das, A.~Banerjee, and Z.~Liu, ``New family of cross {Z}-complementary
	sequences with large {ZCZ} width,'' in \emph{IEEE International Symposium on
		Information Theory (ISIT)}, 2022, pp. 522--527.
	\newpage
	\bibitem{2020PaiCOML}
	C.~{Pai}, S.~{Wu}, and C.~{Chen}, ``Z-complementary pairs with flexible lengths
	from generalized {B}oolean functions,'' \emph{IEEE Commun. Lett.}, vol.~24,
	no.~6, pp. 1183--1187, 2020.
	
	\bibitem{2021LiTSP}
	F.~Li, Y.~Jiang, C.~Du, and X.~Wang, ``Construction of {G}olay complementary
	matrices and its applications to {MIMO} omnidirectional transmission,''
	\emph{IEEE Trans. Signal Process.}, vol.~69, pp. 2100--2113, 2021.
	
	\bibitem{2021WangTIT}
	Z.~Wang, D.~Ma, G.~Gong, and E.~Xue, ``New construction of complementary
	sequence (or array) sets and complete complementary codes,'' \emph{IEEE
		Trans. Inf. Theory}, vol.~67, no.~7, pp. 4902--4928, 2021.
	
	\bibitem{2022PaiTCOM}
	C.~Y. Pai and C.~Y. Chen, ``Two-dimensional {G}olay complementary array
	pairs/sets with bounded row and column sequence {PAPR}s,'' \emph{IEEE Trans.
		Commun.}, vol.~70, no.~6, pp. 3695--3707, 2022.
	
	\bibitem{1980Frank}
	R.~{Frank}, ``Polyphase complementary codes,'' \emph{IEEE Trans. Inf. Theory},
	vol.~26, no.~6, pp. 641--647, 1980.
	
	\bibitem{2021LiuThree}
	C.~Liu, S.~Liu, X.~Lei, A.~R. Adhikary, and Z.~Zhou, ``Three-phase
	{Z}-complementary triads and almost complementary triads,''
	\emph{Cryptography and Communications}, vol.~13, no.~5, pp. 763--773, 2021.
	
	\bibitem{2002Craigen}
	R.~Craigen, W.~Holzmann, and H.~Kharaghani, ``Complex {G}olay sequences:
	structure and applications,'' \emph{Discrete Math.}, vol. 252, no.~1, pp.
	73--89, 2002.
	
	\bibitem{2003ZhangCUWBST}
	C.~Zhang, X.~Lin, and M.~Hatori, ``Novel two dimensional complementary
	sequences in ultra wideband wireless communications,'' in \emph{IEEE
		Conference on Ultra Wideband Systems and Technologies, 2003}, 2003, pp.
	398--402.
	
	\bibitem{2004Turccsany}
	M.~Turcs{\'a}ny and P.~Farka{\v{s}}, ``New {2D-MC-DS-SS-CDMA} techniques based
	on two-dimensional orthogonal complete complementary codes,'' in
	\emph{Multi-Carrier Spread-Spectrum}.\hskip 1em plus 0.5em minus 0.4em\relax
	Springer, 2004, pp. 49--56.
	
	\bibitem{2003Farkas}
	P.~Farkas and M.~Turcsany, ``Two-dimensional orthogonal complete complementary
	codes,'' in \emph{Proc. Joint 1st Workshop on Mobile Future and Symposium on
		Trends in Communications}, Oct. 2003, pp. 21--24.
	
	\bibitem{2006Turcsany}
	M.~Turcsany, P.~Farkas, P.~Duda, and J.~Kralovic, ``Performance evaluation of
	two-dimensional quasi orthogonal complete complementary codes in fading
	channels,'' in \emph{Joint IST Workshop on Mobile Future, 2006 and the
		Symposium on Trends in Communications. SympoTIC}, 2006, pp. 84--87.
	
	\bibitem{2011LiIWSDA}
	Y.~Li and C.~Xu, ``Construction of two-dimensional periodic complementary array
	set with zero-correlation zone,'' in \emph{Proc. Fifth International Workshop
		on Signal Design and Its Applications in Communications}, 2011, pp. 104--107.
	
	\bibitem{2021PaiTIT}
	C.~Y. Pai, Y.~T. Ni, and C.~Y. Chen, ``Two-dimensional binary {Z}-complementary
	array pairs,'' \emph{IEEE Trans. Inf. Theory}, vol.~67, no.~6, pp.
	3892--3904, 2021.
	
	\bibitem{2021AbhishekCOML}
	A.~Roy, P.~Sarkar, and S.~Majhi, ``A direct construction of q-ary {2-D}
	{Z}-complementary array pair based on generalized {B}oolean functions,''
	\emph{IEEE Commun. Lett.}, vol.~25, no.~3, pp. 706--710, 2021.
	
	\bibitem{2021PaiSPL}
	C.~Y. Pai and C.~Y. Chen, ``A novel construction of two-dimensional
	{Z}-complementary array pairs with large zero correlation zone,'' \emph{IEEE
		Signal Process. Lett.}, vol.~28, pp. 1245--1249, 2021.
	
	\bibitem{2020DasTSP2D}
	S.~Das and S.~Majhi, ``Two-dimensional {Z}-complementary array code sets based
	on matrices of generating polynomials,'' \emph{IEEE Trans. Signal Process.},
	vol.~68, pp. 5519--5532, 2020.
	
	\bibitem{2021Zhang}
	\BIBentryALTinterwordspacing
	H.~Zhang, C.~Fan, and S.~Mesnager, ``New constructions of $q$-ary {2-D}
	{Z}-complementary array pairs,'' 2021. [Online]. Available:
	\url{https://arxiv.org/abs/2107.11599}
	\BIBentrySTDinterwordspacing
	
	\bibitem{2014ZLiuYGuan}
	Z.~Liu, Y.~L. Guan, and U.~Parampalli, ``New complete complementary codes for
	peak-to-mean power control in multi-carrier {CDMA},'' \emph{IEEE Trans.
		Commun.}, vol.~62, no.~3, pp. 1105--1113, Mar. 2014.
	
	\bibitem{1999FengShyong}
	K.~Feng, P.~J.~S. Shiue, and Q.~Xiang, ``On aperiodic and periodic
	complementary binary sequences,'' \emph{IEEE Trans. Inf. Theory}, vol.~45,
	no.~1, pp. 296--303, 1999.
	
\end{thebibliography}

\end{document}